\documentclass[fleqn,usenatbib]{mnras}

\usepackage{newtxtext,newtxmath}
\usepackage{xspace}

\usepackage{graphicx}	
\usepackage{amsmath}	
\usepackage{float}

 \title{Virial theorem in clusters of galaxies with MOND}
\author[M. L\'opez-Corredoira et al.]{M. L\'opez-Corredoira,$^{1,2}$\thanks{E-mail: martin@lopez-corredoira.com}
 J. E. Betancort-Rijo$^{1,2}$, 
 R. Scarpa$^{1,2}$,  \v{Z}. Chrob\'akov\'a$^{1,2}$
\\
$^1$ Instituto de Astrof\'\i sica de Canarias, E-38205 La Laguna, Tenerife, Spain\\
$^2$ Departamento de Astrof\'\i sica, Universidad de La Laguna,
E-38206 La Laguna, Tenerife, Spain
}

\date{Last Rev. 25 October 2022}

\begin{document}
\label{firstpage}
\maketitle

\begin{abstract}
A specific modification of Newtonian dynamics known as MOND has been shown to reproduce 
the dynamics of most astrophysical systems at different scales without invoking non-baryonic 
dark matter (DM). There is, however, a long-standing unsolved problem when MOND is applied to 
rich clusters of galaxies in the form of a deficit (by a factor around two) of predicted dynamical mass 
derived from the virial theorem  with respect to observations.
In this article we approach the virial theorem using the velocity dispersion 
of cluster members along the line of sight  rather than 
using the cluster temperature from X-ray data and hydrostatic equilibrium.
Analytical calculations of the virial theorem in clusters for Newtonian gravity+DM and MOND 
are developed, applying pressure (surface) corrections for non-closed systems.
Recent calibrations of DM profiles, baryonic ratio and baryonic ($\beta $ model 
or others) profiles are used, while allowing free parameters to range within the observational constraints.
It is shown that solutions exist for MOND in clusters that give similar results 
to Newton+DM---particularly in the case of an isothermal $\beta $ model for 
$\beta =0.55-0.70$ and core radii $r_c$ between 0.1 and 0.3 times $r_{500}$ (in agreement 
with the known data). The disagreements found in previous studies
seem to be due to the lack of pressure corrections (based on inappropriate hydrostatic equilibrium 
assumptions) and/or inappropriate parameters  for the baryonic matter profiles.
\end{abstract}


   \maketitle
%

\section{Introduction}
\label{.intro}

In present-day astrophysics, many lines of investigation support
the existence of large amounts of non-baryonic dark matter (DM
hereafter) in galaxies and in the Universe at large, the most
obvious example being the asymptotically flat rotation curve of
galaxies, which indicates the existence of massive DM haloes.
Considerable fine tuning is required, however, to justify their
observed properties, the most striking example possibly being
the baryonic Tully--Fisher relation (see e.g. \citet{McG12} and
reference therein). Because of this, over the years more than one
proposal has been made to find alternative explanations not
involving DM.  In particular, it has been shown that a specific
modification of Newtonian dynamics, known as MOND \citep{Mil83a,Mil83b,Mil83c}, is
able to describe many kinds of behaviour of galaxies and
other cosmic structures generally ascribed to the presence of DM.
The basic idea of MOND is that an acceleration ($a_0$) exists, 
below which Newtonian dynamics is no longer valid. 

The MOND hypothesis has profound and far-reaching implications.
Since the seminal papers by \citet{Mil83a,Mil83b,Mil83c}, MOND has been
applied to several astrophysical objects including (in
increasing order of size) wide binary stars \citep{Her12,Her21}, globular
clusters \citep{Sca03,Sca10,Sca11,Her20}, dwarf
galaxies \citep{Mil95,McG13,San21},
gas dominated galaxies \citep{McG12,San19}, spiral
galaxies \citep{San96,Gen11,Mil07} including our Milky Way \citep{Chr20},
elliptical galaxies \citep{Mil03,Dur18,Tia16}, satellites around 
galaxies \citep{Ang08,Kli09}, pairs of galaxies \citep{Mil83c,Sca22}, 
groups of galaxies \citep{Mil19,McG21}, gravitational lenses \citep{San14}, and
cluster of galaxies \citep{San99,San03}. In all cases except one
MOND may describe the observations without the need for
DM. The problematic case being rich
clusters of galaxies, which are a long-standing problem, thus far unsolved 
by MOND, and on which we try to shed some light here.

We know the virial theorem works in clusters of galaxies for
standard Newtonian gravity within the usual assumption of the existence of non-baryonic 
dark matter as predicted by $\Lambda $CDM models 
\citep[e.g.,][]{Evr08,Zha11,Mun13}, but it has not worked for MOND so far.
Using a hydrostactic isothermal model with temperatures derived from X-ray data, 
the MOND mass prediction falls short by a factor $\sim 2$ \citep{San99,Poi05}. 
A more recent analysis by \citet{Ett19} finds that MOND scenarios 
underestimate hydrostatic masses of cluster by 40\% at $r_{1000}$ ($r_x$ being the radius of the sphere for 
which the average density inside it is $x$ times the critical density $\rho _{\rm c}$), 
but with a decreasing tension as the radius increases, and reaches $\sim $15\% at $r_{200}$.
However, this hydrostatic model has certain drawbacks which, according to some authors, may lead to important systematic 
errors of up to a factor 2 for the mass (\citep{Bar96,Bal97}\citep[\S 4.2]{Sad97}).

Other applications of the virial theorem within the famework of MOND are discussed in several works
\citep{Mil94,Mil10,Mil14,Fab09}.
One different method is the application of the virial theorem 
using the velocity dispersion of cluster members along the line of sight.
A study of this kind has been carried out by \citet{Fab09} for the Coma cluster, revealing, within 
Newtonian gravity a mass-to-light ratio M/L$\sim 200$ in solar units (in agreement with 
estimates based on different methods \citep{Car97}), whereas for MOND it is three times lower 
(still problematic for MOND). Instead of using optical surface brightness to trace the baryonic 
mass (and non-baryonic mass for Newtonian gravity), 
a derivation of baryonic mass calibrated with X-ray data would be more
accurate.

Here, by allowing free parameters to range within the observational constraints,
 we revisit the application of the virial theorem in clusters of galaxies 
using the velocity dispersion of cluster members along the line
of sight. Recent calibrations of DM profiles, baryonic ratio and baryonic (the $\beta $ model or 
others) profiles are used. 
We also apply pressure corrections for non-closed systems (usually overlooked in the literature).
Both Newtonian gravity and MOND are considered, 
although we pay special interest to the latter case.

\section{Application of the virial theorem}

\subsection{Virial theorem}

We assume spherical symmetry in a rich cluster with mass density distribution $\rho (r)$ and
mass interior to each radius $r$ 
\begin{equation}
M(r)\equiv 4\pi \int _0^r dx\,x^2\rho (x)
.\end{equation}

The potential energy with MOND or Newtonian gravity within a radius $r_{\rm max}$ is \citep{Fab09}
\begin{equation}
V(r_{\rm max})=-4\pi G\int_0^{r_{\rm max}}dr\,r\sqrt{1+\left(\frac{r}{r_{cM}(r)}\right)^2} \rho (r)M(r)
,\end{equation}
with $r_{cM}(r)\longrightarrow \infty$ for Newtonian gravity. The distance $r_{cM}(r)$ is related to the
usual parameter $a_0$ by means of
\begin{equation}
r_{cM}(r)=\sqrt{\frac{G \,M(r)}{a_0}}
\end{equation}
and $a_0=1.2\times 10^{-10}$ m/s$^2$ for MOND or $a_0=0$ for Newton.

The kinetic energy is 
\begin{equation}
K(r_{\rm max})=3\pi \int _0^{r_{\rm max}}dR\,R\,\sigma_v^2(R)\Sigma(R)
,\end{equation}\begin{equation}
\Sigma (R)=2\int _{0}^{+\sqrt{r_{\rm max}^2-R^2}}dz\,\rho(\sqrt{R^2+z^2})
,\end{equation}
where $\sigma _v(R)$ is the line-of-sight velocity dispersion (in the rest-frame of the cluster) as a function
of the projected distance $R$, 
and $\Sigma (R)$ is the surface mass density. This can also be expressed as:
\begin{equation}
K(r_{\rm max})=\frac{3}{2} M(r_{\rm max}) \sigma _{v,r<r_{\rm max}}^2
.\end{equation}

In a virialized cluster, in the limit of $r_{\rm max}\rightarrow \infty$, 
the following condition should be followed:
\begin{equation}
\label{virial}
2K(r_{\rm max})+V(r_{\rm max})=0
.\end{equation}
Hence,
\begin{equation}
\sigma _{v.r<r_{\rm max}}=\sqrt{\frac{-V(r_{\rm max})}{3M(r_{\rm max})}}
.\end{equation}

\subsection{Corrections to the virial theorem that include the pressure term}

None of the above considerations can be exact unless we make $r_{\rm max}=\infty $, which
would make the cluster a closed system. For a finite virial radius $r_{\rm max}$ there is a pressure
term due to exchange of galaxies and other types of matter between the sphere 
within radius $r_{\rm max}$ and the space beyond it. 
The relevance of the surface (pressure) term in standard gravity is known \citep[e.g.,][]{The86,Car96,Car97,Gir98}.

When the pressure term is included,
the expression of the dispersion of velocities would be (see appendix \ref{.pressure}):
\begin{equation}
\sigma _{v,r<r_{\rm max}}^2= \sigma _{v,r<r_{\rm max}, P=0}^2
\end{equation}\[
+
\frac{4\pi }{(3-2\beta _a )} r_{\rm max}^3\rho (r_{\rm max}) \sigma _{v,r=r_{\rm max}}^2\frac{1}{M(r_{max})} 
,\]
where $\beta _a$ is the velocity anisotropy parameter: $\beta _a=1-\frac{\sigma _\theta ^2}{\sigma _r^2}$. 
The second term is associated with the pressure. Assuming $\beta _a =1/4$ \citep{Kli09} and
neglecting the variation of $\sigma _v$ with the
radius, we get for the virial radius $r_{\rm max}=r_{200}$ (which by definition follows
$200\rho _c=\frac{M(r_{200})}{\frac{4}{3}\pi r_{200}^3}$, with $\rho _c$ the critical density) 
\begin{equation}
\label{fp}
\sigma _{v,r<r_{200}}\approx \sigma _{v,r<r_{200}, P=0}\times F_P,\end{equation}\[
F_P=\left[1-\frac{6}{5}\frac{\rho (r_{200})}{200\, \rho _c}\right]^{-1/2}
.\]
Note that both $\rho $ and
$\rho _c$ should refer to the same matter, including either  non-baryonic dark matter (in Newton+DM)
or only baryonic (in MOND).
This corrective factor by pressure, $F_P$, depends on the profile.

\subsection{Newtonian gravity}

For standard Newtonian gravity a good description of the hydrostatic mass profile of clusters
as derived from X-ray data is obtained with NFW profiles \citep{Nav97} with scale $r_{200}$ \citep{Ett19}. 

Using $\rho (r)$, $M(r)$ of NFW profiles
(formulae of appendix \S \ref{.NFW}, with concentration index $C$) 
and $a_0=0$, we get  potential energy 

\begin{equation}
V(r_{\rm max})=-\frac{4\pi G\rho _0}{C\left[\ln (1+C)-\frac{C}{1+C}\right]}r_{200}^2\,M_{200} 
\end{equation}\[
\times
\int_0^{\frac{r_{\rm max}}{r_{200}}} dx \frac{[\ln (1+C\,x)-\frac{C\,x}{1+C\,x}]}{(1+C\,x)^2}
.\]
Taking the values of $r_{200}$ and $\rho _0$ from Eq. (\ref{mr}):
\begin{equation}
V(r_{\rm max})
=-(8.707\times 10^{55}\ {\rm J})\frac{C^2}{\left[\ln (1+C)-\frac{C}{1+C}\right]^2}\left(\frac{M_{200}}{10^{14}\ {\rm M_\odot }}\right)^{5/3}
\end{equation}\[
\times \int_0^{\frac{r_{\rm max}}{r_{200}}} dx \frac{[\ln (1+C\,x)-\frac{C\,x}{1+C\,x}]}{(1+C\,x)^2}.\]

The kinetic energy is
\begin{equation}
K(r_{\rm max}) 
=(2.985\times 10^{50}\ {\rm J})[\sigma _{v,r_{\rm max}}({\rm km/s})]^2\left(\frac{M_{200}}{10^{14}\ {\rm M_\odot }}\right)
\end{equation}\[
\times \frac{\left[\ln \left(1+\frac{C\,r_{\rm max}}{r_{200}}\right)-\frac{C\,r_{\rm max}}{r_{200}+C\,r_{\rm max}}\right]}
{\left[\ln (1+C)-\frac{C}{1+C}\right]}
.\]

For the virial radius $r_{\rm max}=r_{200}$, and
applying the virial theorem Eq. (\ref{virial}), with the pressure correction
referred at Eq. (\ref{fp}) ($F_P$ is independent of the mass), we get
\begin{equation}
\sigma _{v,r200,{\rm Newton+DM}}=(382 \ {\rm km\ s^{-1}})\left(\frac{M_{200}}{10^{14}\ {\rm M_\odot }}\right)^{1/3}
\end{equation}\[\times
\frac{C}{\left[\ln (1+C)-\frac{C}{1+C}\right]} 
\left[\int_0^{1} dx \frac{[\ln (1+C\,x)-\frac{C\,x}{1+C\,x}]}{(1+C\,x)^2}\right]^{1/2}\times F_P(C)
,\]\[
F_P=\left[1-0.400\frac{C^2}{(1+C)^2\left[\ln (1+C)-\frac{C}{1+C}\right]}\right]^{-1/2}
.\]
This dependence with the mass to the power of 1/3 is also well known from simulations \citep{Evr08,Mun13}.

\begin{figure}
\vspace{0cm}
\centering
\includegraphics[width=8.5cm]{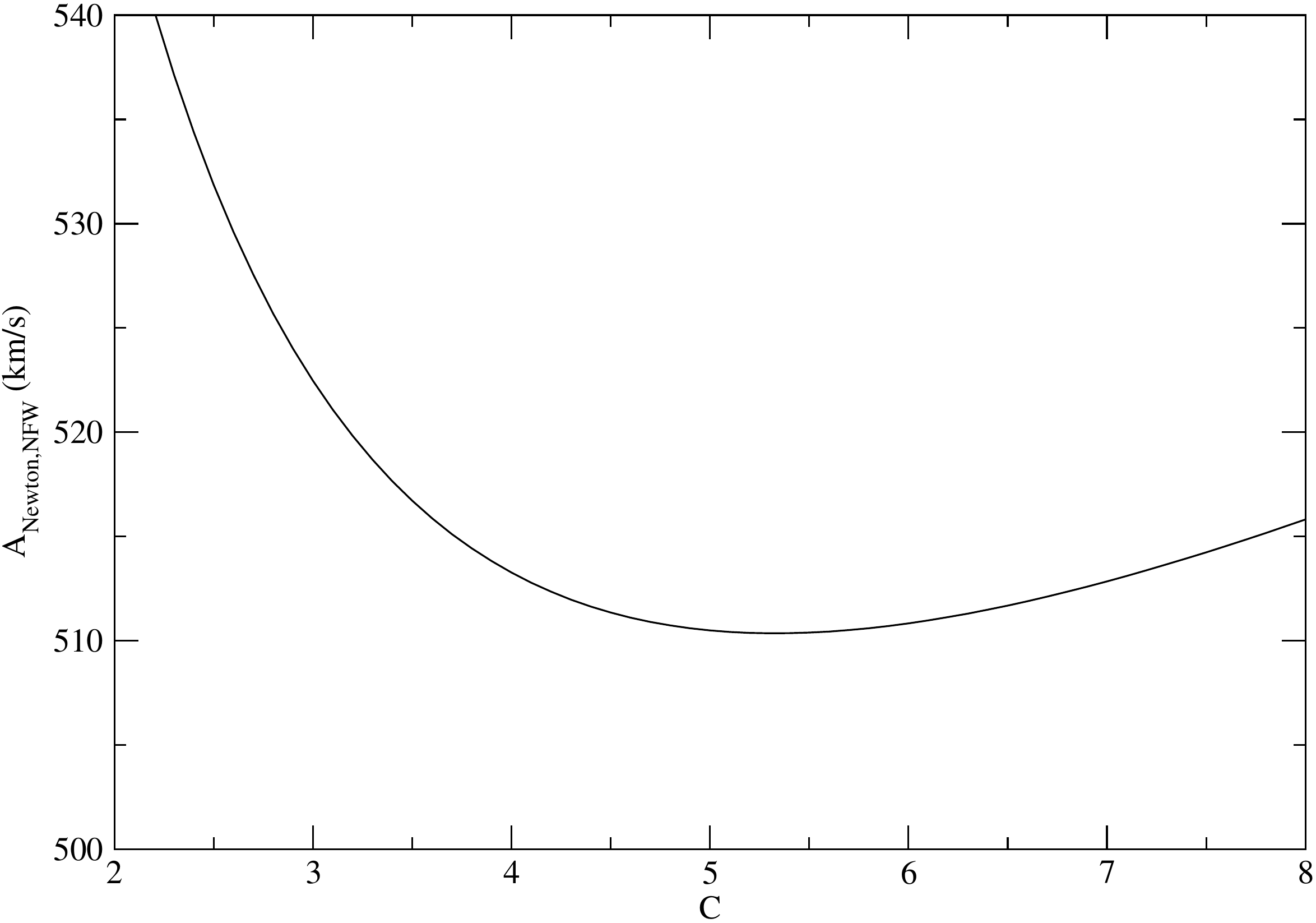}
\caption{Value of $A_{\rm Newton,NFW}(C)$ in Eq. (\ref{sigmavn}) as a function of $C$.}
\label{Fig:ANFW}
\end{figure}

Throughout this paper, we shall calculate the dispersion of velocities 
as a function of $M_{500}\equiv M(r_{500})$. This amount might also be related to other parameters or measurements; for instance,
the Sunyaev--Zel'dovich effect amplitude ($Y_{\rm SZ}$) \citep{Arn10,Lop17,Agu21}.
In terms of $M_{500}$ [with $M_{200}/M_{500}$ derived from Eq. (\ref{mr})], and including all the
dependence of $C$ in a single factor,
\begin{equation}
\label{sigmavn}
\sigma _{v,r200,{\rm Newton+DM}}=A_{\rm Newton,NFW}(C)
\left(\frac{M_{500}}{10^{14}\ {\rm M_\odot }}\right)^{1/3}
,\end{equation}
where $A_{\rm Newton,NFW}(C)$ is plotted in Figure \ref{Fig:ANFW}. The dependence on $C$ is quite small
for $C$ between 2 and 8. Several values are given in the literature: 
from $C=2.9\pm 0.2$ \citep{Lin04,Mac08} derived from
analyses of observational X-ray data; or $C=4.6^{+1.8}_{-1.1}$ from purely theoretical dynamical
 models in \citet{Pra12} (the error bars represent here the r.m.s., not the error of the average; 
assuming that the 10, 90 per cent percentiles of fig. 13 in \citet{Pra12} are 1.28 times the 
r.m.s., as it corresponds to a Gaussian distribution). 
The concentration index $C$ has a modest dependence on mass \citep{Lin04,Mac08,Pra12,Ett19}.
The above range $2.7<C<6.4$ gives
a variation of only $\sim 2$\% of $A_{Newton,NFW}(C)$, which is
negligible compared to other sources of errors. We take in the following as default $C=3$, for which 
$A_{\rm Newton,NFW}(C)=522$ km s$^{-1}$ ($F_P=1.244$).

\subsection{MOND with only baryonic matter}

We set $a_0=1.2\times 10^{-10}$ m s$^{-2}$.
For MOND, there is only baryonic mass.
To calculate the amount of baryonic mass, we use the relationship obtained 
by \citet{Gon13}:
\begin{equation}
\label{mbar}
M_{\rm bar.500}(M_{500})=(0.117\pm 0.004)\times 10^{14}\times \left(\frac{M_{500}}{10^{14}\ 
{\rm M_\odot }}\right)^{1.16\pm 0.04}\ {\rm M_\odot }
.\end{equation}

For a baryonic density distribution of the type $\rho _{\rm bar.}(r)=\rho _0J\left(\frac{r}{r_c}\right)$, 
with $J$ a generic function 
and cluster core radius $r_c$ proportional to $r_{500}$ [we define the parameter independent of the mass 
$x_{500}\equiv \frac{r_c}{r_{500}}$; \citep{Pac16}], the potential and kinetic energies are

\begin{equation}
\label{VMOND}
V(r_{\rm max})=
-(6.366\times 10
^{55}\ {\rm J})\frac{1}{x_{500}I^2(r_{500})}\left(\frac{M_{\rm bar.500}}{10^{14}\ {\rm M_\odot }}\right)^{5/3}
\end{equation}\[
\times
\int_0^{\frac{r_{\rm max}}{r_c}} dx \,x\,J(x)\, I(x\,r_c)\,R(x)\]
,\[
R(x)=\sqrt{1+15.57\,I(r_{500})x_{500}^2\left(\frac{M_{\rm bar.500}}{10^{14}\ {\rm M_\odot }}\right)^{-1/3}
\frac{x^2}{I(x\,r_c)} }
,\]\[
I(r)=\int _0^{r/r_c}dx\,x^2J(x),\]

\begin{equation}
\label{KMOND}
K(r_{\rm max})=(2.980\times 10^{50}\ {\rm J})[\sigma _{v,r_{\rm max}}({\rm km/s})]^2 
\left(\frac{M_{\rm bar.500}}{10^{14}\ {\rm M_\odot }}\right)
\frac{I(r_{max})}
{I(r_{500})}
.\end{equation}

\begin{figure}
\vspace{0cm}
\centering
\includegraphics[width=8.5cm]{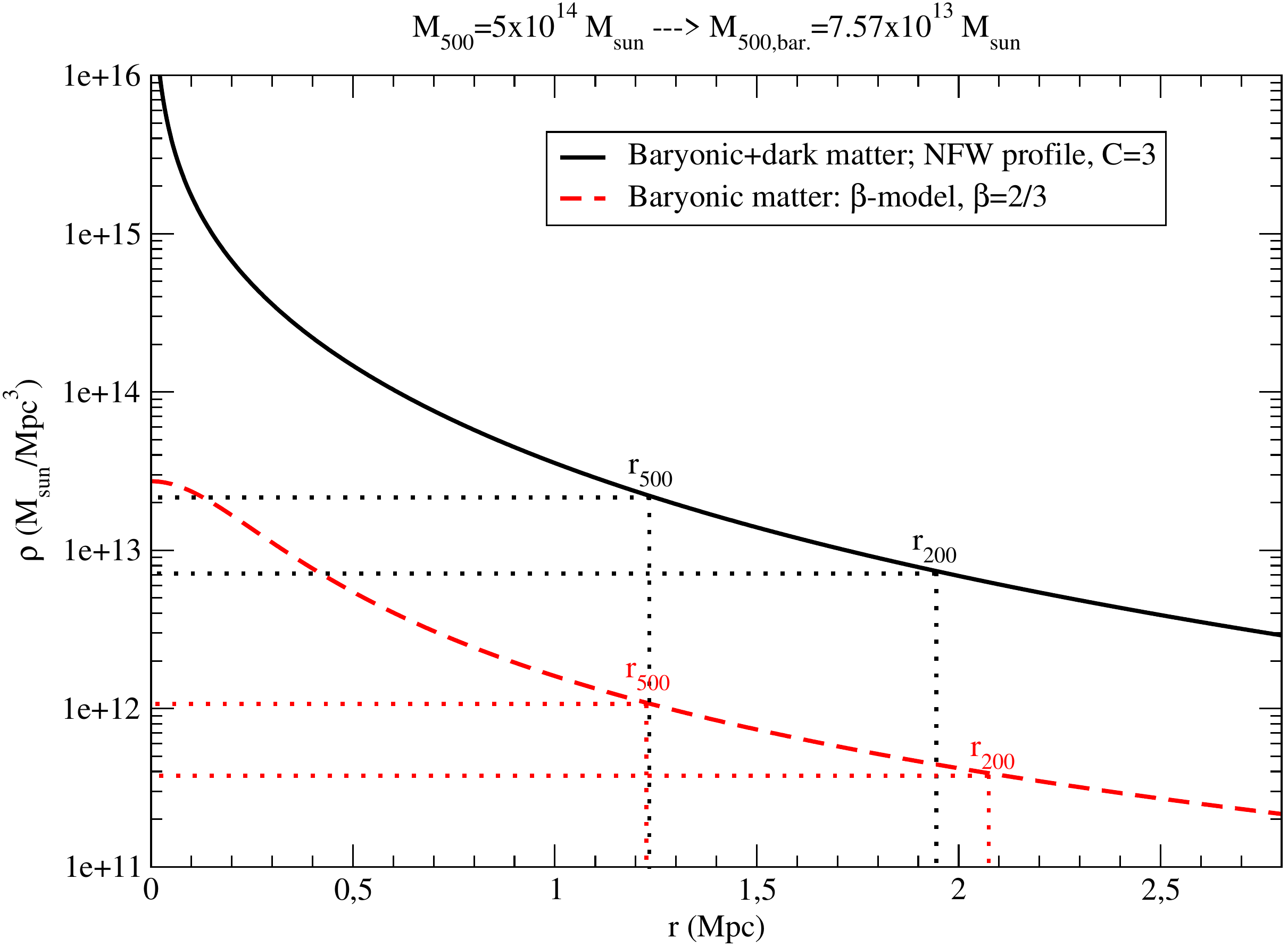}
\caption{Numerical example of profiles with NFW model with $C=3$ for baryonic+dark matter and $\beta $ model
with $\beta =2/3$ for baryonic matter, for a mass $M_{500}=5\times 10^{14}$ M$_\odot $, equivalent to 
$M_{\rm bar.500}=0.773\times 10^{14}$ M$_\odot $. In dotted lines, the respective positions of $r_{500}$, 
$r_{200}$, $\rho (r_{500})$, $\rho (r_{200})$; where $r_x$ is the radius of the sphere for 
which the average density inside it is $x$ times a critical density $\rho _c=8.5\times 10^{-27}$ kg m$^{-3}$ for the case
with dark matter or $\rho _{cb}=1.33\times 10^{-27}$ kg m$^{-3}$ with only baryonic density.
Note the similarity of $r_{500}$ for both distributions.}
\label{Fig:profiles}
\end{figure}

Note that the radiii $r_{500}$ or $r_{200}$ are here quite similar to the one obtained from Newton+dark 
matter(DM)+NFW. For instance,
with Eqs. (\ref{mbar}), (\ref{mrb}), we get
\begin{equation}
\label{r500MOND}
r_{500}=
1.345\ {\rm Mpc}\times \left(\frac{M_{\rm bar.500}}{10^{14}\ {\rm M_\odot }}\right)^{1/3}
\end{equation}\[
=0.658\ {\rm Mpc}\times \left(\frac{M_{500}}{10^{14}\ {\rm M_\odot }}\right)^{0.386}
,\]
which is similar although slightly lower than the $r_{500}$ from Eq. (\ref{mr}) of Newton+DM.
In Fig. \ref{Fig:profiles}, we offer a plot with a numerical example for $M_{500}=5\times 10^{14}$ M$_\odot $.
This approximate coincidence is expected because the ratio of baryonic/total matter in the cluster is similar
to the ratio of baryonic/total matter in the Universe ($\Omega _b/\Omega_m$ and 1 respectively
for Newton+DM and MOND).
The fact that this ratio is the same one in clusters and in the Universe implies that the  central 
cluster density in MOND is similar to the cluster central density in standard gravity times $\Omega _b/\Omega_m$, 
thus leading to similar $r_{500}$'s in the two models.
The fact that the similarity is tighter for $r_{500}$ than for $r_{200}$ is a coincidence.

For $r_{\rm max}=r_{200}$, and only baryonic density (we take critical baryonic density 
$\rho _{cb}=1.33\times 10^{-27}$ kg m$^{-3}$; see appendix \ref{.beta}). the correction of the virial theorem due the pressure is a factor
\begin{equation}
\label{fpMOND}
F_P=\left[1-\frac{3}{(3-2\beta _a)}\,\frac{5J\left(\frac{r_{200}}{x_{500}r_{500}}\right)}
{6x_{500}^3I(r_{500})} \right]^{-1/2}
,\end{equation}
and we assume, as previously, $\beta _a=1/4$; except in the extreme case of very low-concentrated
(almost flat) profiles with very high densities at $r_{200}$ 
[where $R_2\equiv \frac{\rho _{\rm bar.}(r_{200})}{200\rho _{cb}}= \frac{5J\left(\frac{r_{200}}{x_{500}r_{500}}\right)}
{6x_{500}^3I(r_{500})}$ is larger than 0.75], for which we assume a $\beta _a$ between 0 and 0.25, in a linear
dependence with $R_2$ ($\beta _a=1-R_2$), in order to avoid
a negative root square.

\subsubsection{$\beta $ isothermal model}

We now assume a $\beta $ model for the baryonic matter profile, which
is usually adopted to fit the intracluster gas distribution \citep{Arn09}:
\begin{equation}
J(x)=\frac{1}{(1+x^2)^{1.5\beta }}
,\end{equation}
where $x=r/r_c$.

\begin{figure*}
\vspace{0cm}
\centering
\includegraphics[width=8.5cm]{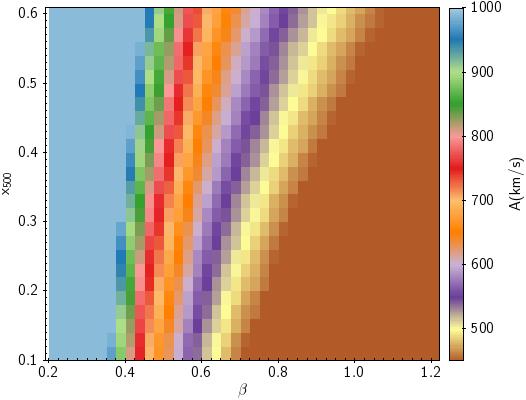}
\hspace{.2cm}
\includegraphics[width=8.5cm]{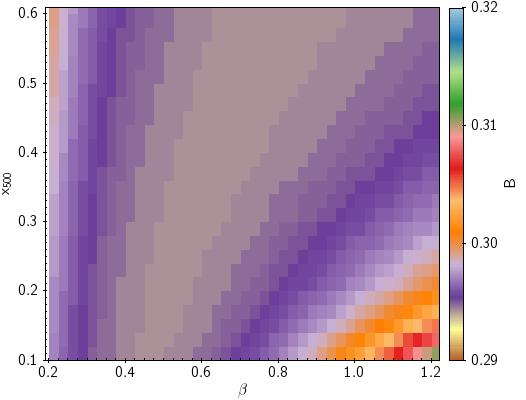}
\caption{Values of $A$, $B$ in Eq. (\ref{sigmaMOND}) as a function of $\beta $ and $x_{500}$ such
that the cluster core radius $r_c=x_{500}r_{500}$ in a $\beta $ model of the baryonic matter+MOND.}
\label{Fig:AB}
\end{figure*}

Following the formulae of appendix \ref{.beta},
for $r_{\rm max}=r_{200}$,
applying the virial theorem Eq. (\ref{virial}), 
with the pressure correction
referred at Eq. (\ref{fp}) [we obtain $F_P$ given by Eq. (\ref{fpMOND}), independent of the mass],
we get a dependence that is fitted  in the range $M_{500}=(1-10)\times 10^{14}$ M$_\odot $ with high accuracy by
\begin{equation}
\label{sigmaMOND}
\sigma _{v,r_{200},{\rm MOND}}\approx A(\beta, x_{500})
\left(\frac{M_{500}}{10^{14}\ {\rm M_\odot }}\right)^{B(\beta, x_{500})}
,\end{equation}
where $A(\beta, x_{500})$ and $B(\beta, x_{500})$ are plotted at Fig. \ref{Fig:AB}.
The exponent $B$ is almost constant, between 0.29 and 0.30 for most of the cases. The amplitude $A$
is however quite dependent on the parameters of the $\beta $ model.

For the usual values of $\beta=2/3$, $x_{500}=0.15$ \citep{Pac16}, 
$A=493$ km s$^{-1}$, $B=0.295$. The pressure factor correction is $F_P=1.357$.
For comparison, the prediction of the velocity dispersion with only baryonic matter
following $\beta $-model profile and Newtonian gravitation [equivalent to substituting
15.57 for 0 within the root square inside the integral of Eq. (\ref{VMOND})], for the same
parameters $\beta=2/3$, $x_{500}=0.15$ and the same range of masses, is $A=206$ km s$^{-1}$,
$B=0.387$. Clearly, the effect of MOND is quite significant. 
It is almost enough to compensate for the absence of non-baryonic dark matter: it is 5--15\%
 (depending on the mass) lower than $\sigma_v$ for Newton+DM for $C=3.0$.

For other values of the parameters $\beta $, $x_{500}$ can also approximately reproduce the
 dynamics of Newton+DM. Values of $A$ similar to the amplitude of Newton+DM (=522 km s$^{-1}$) 
are obtained in the yellow--violet colour area
of the left plot of Fig. \ref{Fig:AB}: $\beta $ between 0.55 and 0.80. 
For reasonable values of $0.1<x_{500}<0.3$ [which lead, through Eq. (\ref{r500MOND})
to $0.12<r_c<0.36$ Mpc for an average mass of $M_{500}=5\times 10^{14}$ M$_\odot $; of the order of $r_c=0.25$ Mpc 
given by \citet{Jon84}], the values of $\beta $ are constrained between 0.55 and 0.70 in order
to match the Newtonian amplitude. In Fig. \ref{Fig:sigmav}, we plot the dispersion of velocities for
the parameters $\beta =0.65$, $x_{500}=0.3$ which are very close to the Newton+DM results, giving $A=553$ km s$^{-1}$, $B=0.294$,
$F_P=1.448$ (note that the pressure factor here  is 16\% higher than with Newton).
The values of $\beta $ in the literature \citep[e.g.,][]{Bah94,Hen09} are of the same order,
between 0.50 and 0.65 for rich clusters.

\subsubsection{The \citet{Pat15} model}

The isothermal $\beta $-model is known to be insufficient for characterizing the range of cluster gas distributions
\citep{Vik06,Pat15}. Other profiles could be used that give a better fit to the gas distribution. 
Here we use the one given by \citet{Pat15}:
\begin{equation}
\rho _{\rm bar.}(r)=\Gamma f_g \left(\frac{r}{s}\right)^{3\Gamma -3}\rho _{\rm DM}\left[s\left(
\frac{r}{s}\right)^\Gamma \right]
,\end{equation}
where $f_g$ is the fraction of gas with respect the total (in Newton+DM), i.e. 
$f_g=\frac{M_{\rm bar.500}}{M_{500}}$; $\rho _{\rm DM}(r)$ is the profile of the 
total mass including dark matter, in our case given by the NFW profile (see \S \ref{.NFW}) with
scale $r_s$ ($=\frac{r_{200}}{C}$, with concentration index $C$), 
and $\Gamma $, $s$ are two extra free parameters. For $\Gamma =1$, we would have that the baryonic
mass traces the dark matter [$\rho _{\rm bar.}(r)=f_g\rho _{\rm DM}(r)$].
Like the $\beta $ model, the above expression is also motivated on a theoretical basis
 within standard Newtonian gravity. Here, with MOND, we use it because it simply 
fits the observational profile of gas in clusters of galaxies, as a function that describes 
baryonic matter, and the theoretical derivation would have no sense.

This gives a 
\begin{equation}
J(x)=\frac{x^{2\Gamma -3}}{(1+f_sx^\Gamma )^2}
,\end{equation}\[
f_s=\frac{s}{r_{s}}, \ \ r_c=s,
\]
and applying the virial theorem from Eqs. (\ref{virial}), with $V$ and $K$ of Eqs. (\ref{VMOND}), (\ref{KMOND}),
we get a dependence that is fitted  in the range $M_{500}=(1-10)\times 10^{14}$ M$_\odot $ with high accuracy by
\begin{equation}
\label{sigmaMOND2}
\sigma _{v,r_{200},{\rm MOND}} \approx E(\Gamma, x_{500},f_s)
\left(\frac{M_{500}}{10^{14}\ {\rm M_\odot }}\right)^{D(\Gamma, x_{500},f_s)}
.\end{equation}

\begin{figure*}
\vspace{0cm}
\centering
\includegraphics[width=8.5cm]{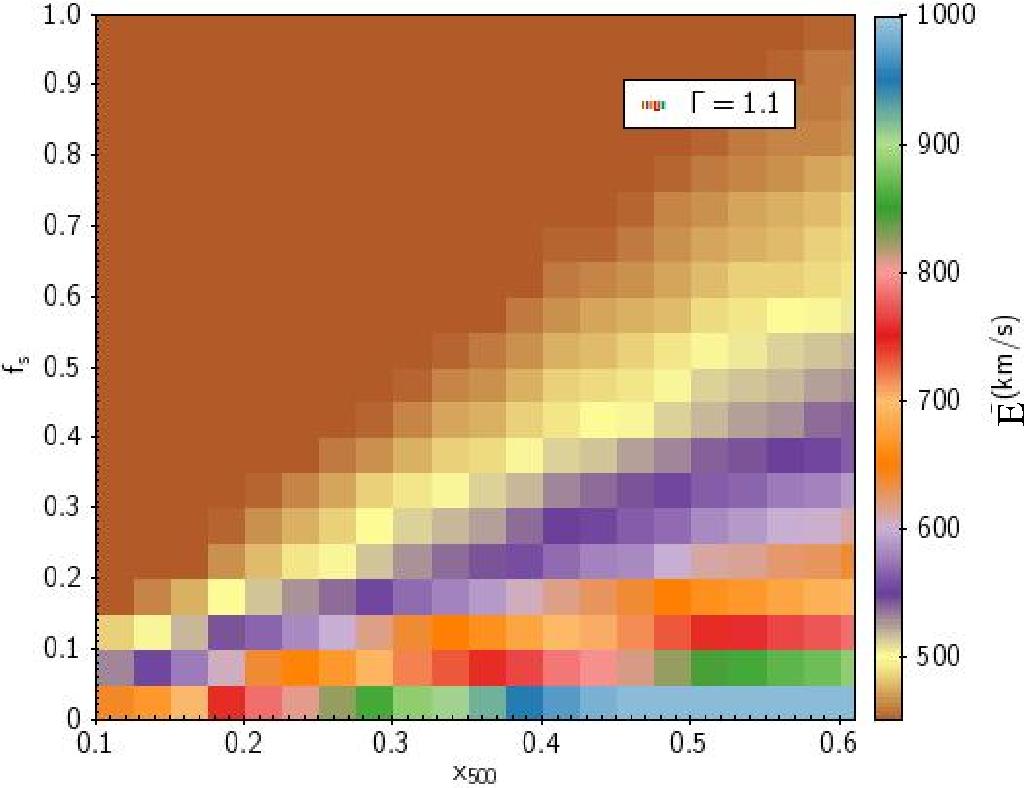}
\hspace{.2cm}
\includegraphics[width=8.5cm]{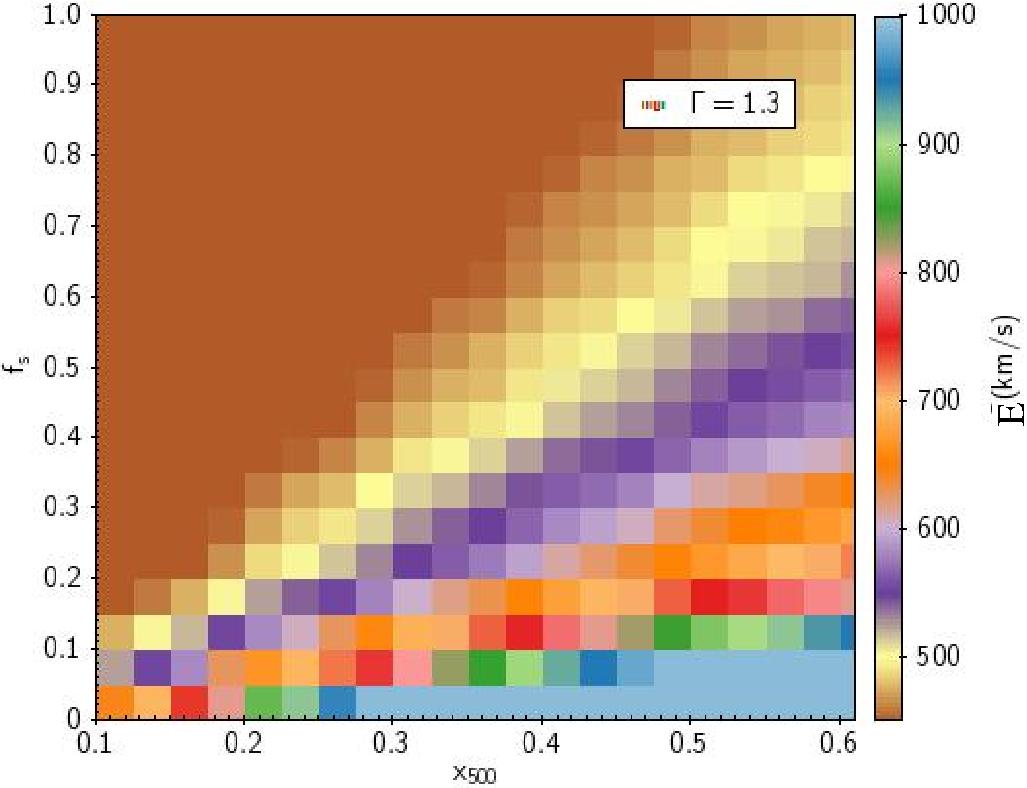}
\hspace{.2cm}
\includegraphics[width=8.5cm]{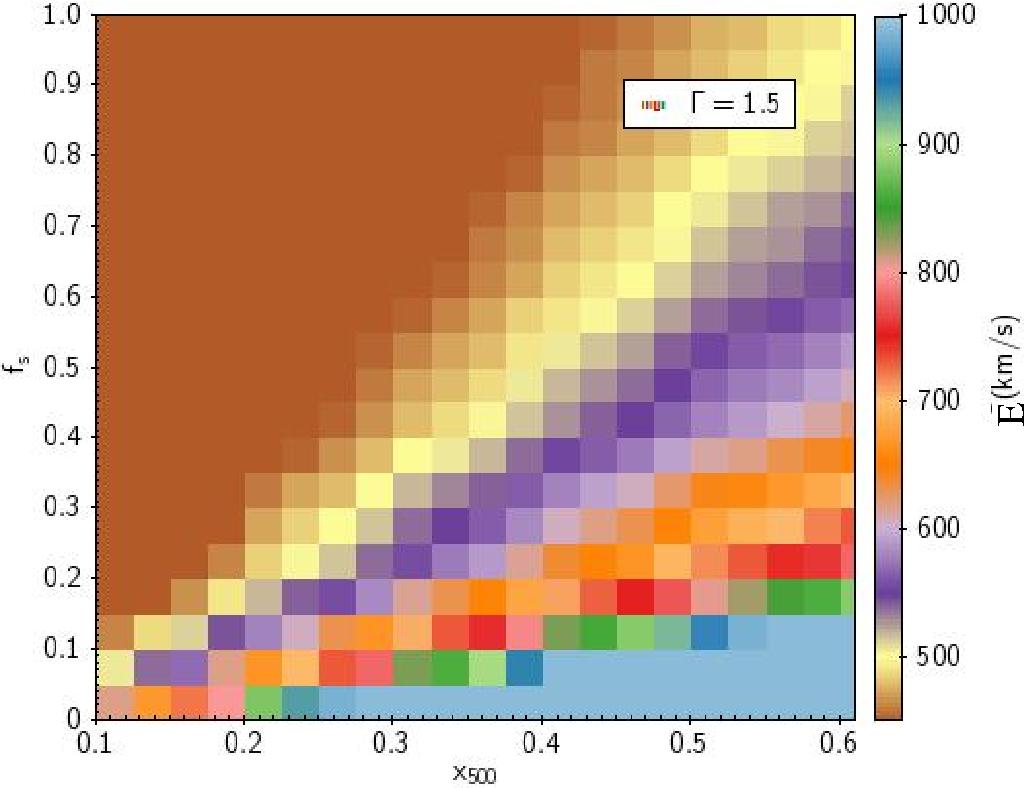}
\caption{Values of $E$ in Eq. (\ref{sigmaMOND2}) as a function of $\Gamma $, $x_{500}$ and $f_s$ 
in the \citet{Pat15} model of the baryonic matter+MOND.}
\label{Fig:CD}
\end{figure*}

For the range $1<\Gamma \le 2$, $0<x_{500}\le 1$, $0<f_s\le 0.4$,
the exponent $D(\Gamma, x_{500},f_s)$ falls always in the range between 0.29 and 0.34
in approximate agreement with Newton+DM.
The amplitude $E(\Gamma, x_{500},f_s)$ is plotted at Fig. \ref{Fig:CD}
for $\Gamma =1.1$, 1.3, 1.5, which are within the constraints obtained by \citet{Pat15}
of $1<\Gamma \le 1.5$. There is a wide range of possible values compatible with Newton+DM.
For instance, if we assume an average value of $\Gamma =1.5$, $x_{500}=0.5$ (hence, $s=0.5r_{500}$), 
$E=646$ km s$^{-1}$ implies $f_s\approx 0.6$ (hence, $r_s\sim 0.8\,r_{500}$, equivalent to a 
concentration index $C\sim 2$). 
 
\subsection{Comparison with observations}

\begin{figure*}
\vspace{0cm}
\centering
\includegraphics[width=15cm]{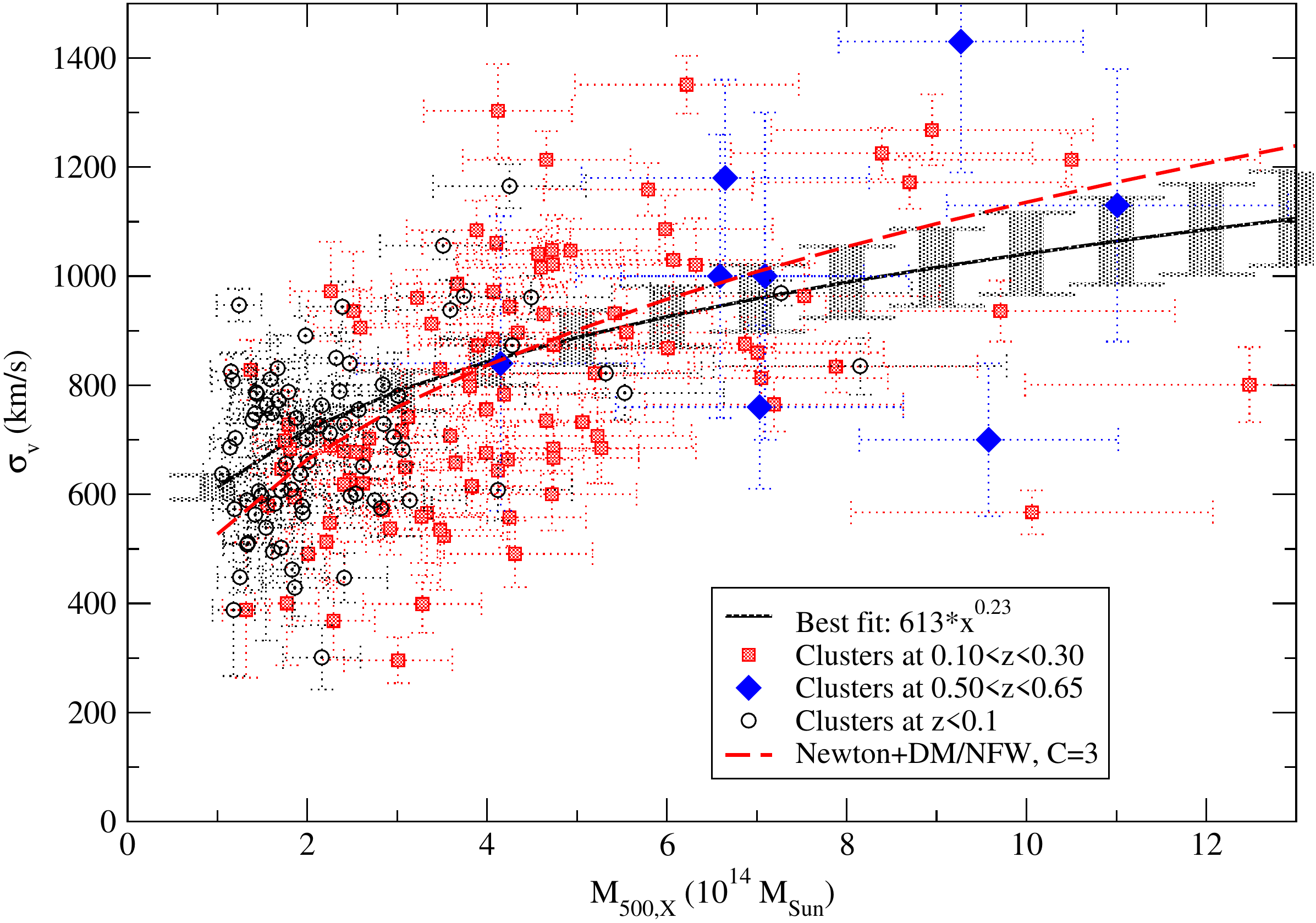}
\hspace{.2cm}
\includegraphics[width=15cm]{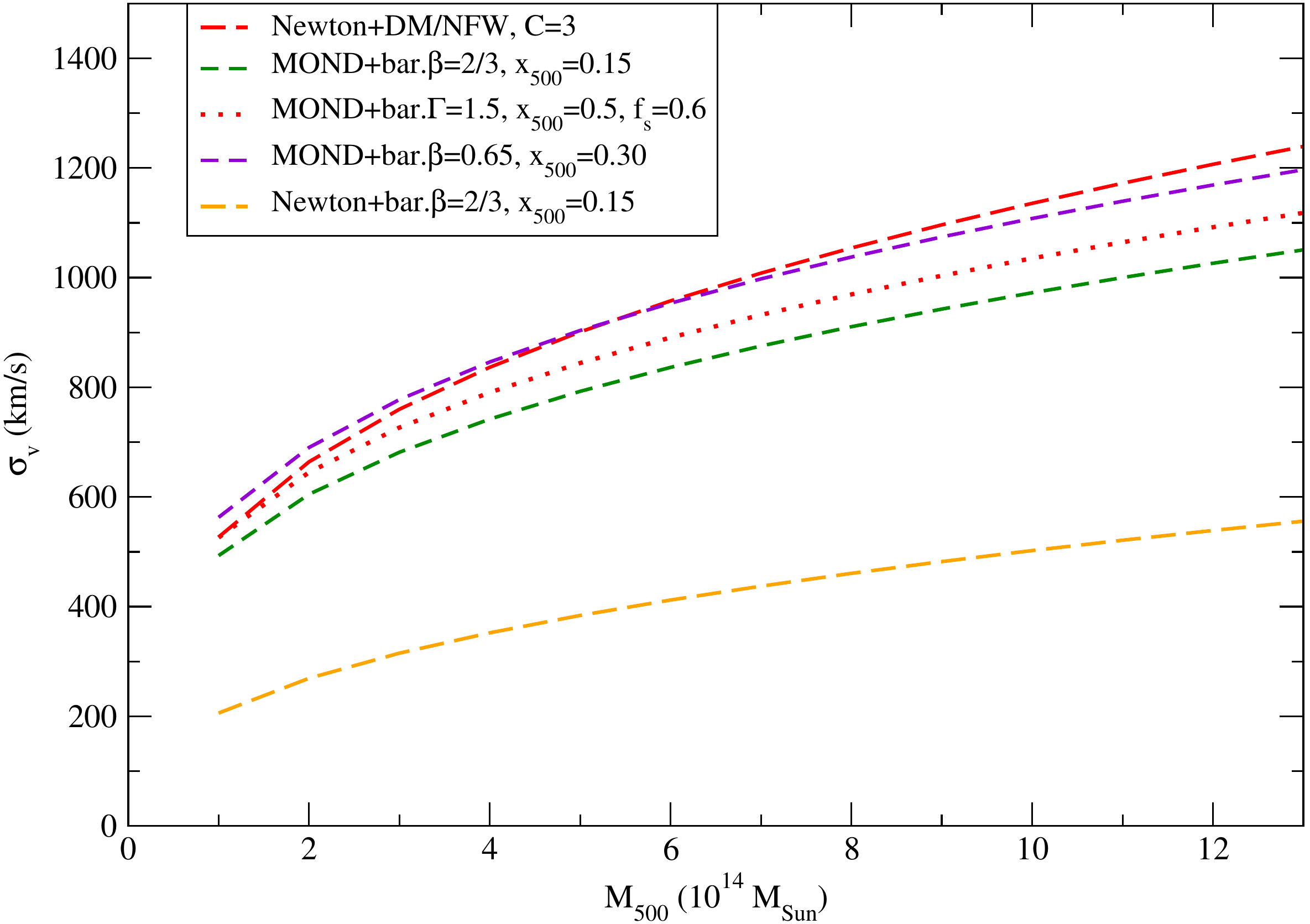}
\caption{Upper panel: Rest-frame dispersion of velocites along the line of sight 
as a function of the $M_{500}$ mass in 178 clusters of galaxies. 
$M_{500}$ represents the total mass within the radius $r_{500}$ for Newtonian gravity+dark matter,
whereas in MOND it is related to the total (baryonic) mass within $r_{500}$ through
$M_{\rm bar.500}=0.117\times 10^{14}\times \left(\frac{M_{500}}{10^{14}\ {\rm M_\odot }}\right)^{1.16}\ {\rm M_\odot }$. 
The lines represent the power-law fit (the shaded area covers the range within a 1-$\sigma $
error of the fit), or the predictions with the
standard Newtonian gravity+non-baryonic dark matter following a NFW profile.
Bottom panel: predictions with the
standard Newtonian gravity+non-baryonic dark matter following a NFW profile or MOND for different 
baryonic matter profiles, or Newtonian gravity with only baryonic matter.}
\label{Fig:sigmav}
\end{figure*}

Estimates of masses  and velocity dispersions carried out by other teams
for some clusters are shown in Figure \ref{Fig:sigmav}.

At low ($z<0.10$) and intermediate ($0.10<z<0.30$) redshifts, we use  velocity dispersion data 
within $r_{200}$ from clusters of \citet[Table 2]{Soh20}, including the mass $M_{500}$ within this table 
estimated from X-ray observations by \citet{Pif11}. We use only the clusters with $M_{500}\ge 10^{14}$ M$_\odot $.
These comprise 74 clusters with $z<0.10$ and 96 clusters with $0.10<z<0.30$. The $M_{500}$ errors are
not provided; here we assume they have a 20\% of error, which is typical of other estimates of X-ray masses
\citep{Vik06,Wal12,Mar14,Hai18,Whe21}.

X-ray data for $M_{500}$, $\sigma _v$, 
Error$(\sigma _v$ are obtained for eight high-redshift clusters ($0.50<z<0.65$)
from the NIKA2 cluster survey \citep{May19}: see Table \ref{Tab:clusters}. Rest-frame velocity dispersions 
were calculate using public Sloan Digital Sky Survey (SDSS) data
of galaxies' velocities and applying a biweight technique \citep{Bee90}. 
X-ray masses were derived from REFLEX \citep{Boh04} and REXCESS \citep{Boh07} cluster 
surveys applying the method by \citet{Arn10}, whose error bars are estimated with the relative
error bar of the X-ray luminosities (when available, or the average value of similar clusters
of this sample otherwise).

Note that the values of $M_{500}$ from X-ray data correspond always to the estimations using the
standard model Newton+DM. As it was remarked throughout the paper, in
a MOND model, it would be approximately related to the total (baryonic) mass within $r_{500}$ through
$M_{\rm bar.500}=0.117\times 10^{14}\times \left(\frac{M_{500}}{10^{14}\ {\rm M_\odot }}\right)^{1.16}\ {\rm M_\odot }$.

In cases with small numbers of galaxies, there may be some important
biases in the galaxy cluster velocity dispersion \citep{Fer20}. Here we do not introduce any correction
to take them into account, since the number of galaxies per cluster is high enough and the
corrections of the statistics for small numbers are negligible.
For the comparison with the theoretical predictions, we also assume that the r.m.s. of $\sigma _v$ is 
much smaller than its average value within $r_{200}$, as is usually the case \citep{Fer20}.
The observed velocities may be slightly different from the average because
the average line-of-sight velocities were measured within a radius smaller than $r_{200}$
and our approximation of almost constant dispersion of velocities with radius might introduce some higher values
of dispersion in the observations than in the theory. We assume that these differences are lower than the error bars. 

Other effects could produce a few small systematics \citep{Kri14}: relativistic effects of high 
velocities, gravitational redshift, and gravitational lensing in a curved space, which 
would decrease the Hubble--Lema\^itre parameter, intergalactic baryonic matter, gravitational aberration, etc.

\begin{table*}
\caption{Clusters at high redshift ($z>0.5$ used for our analysis, ordered by increasing redshift.
The third column indicates the number of galaxies used for the measurent of 
the rest-frame velocity dispersion (column 5). Column 4 gives the estimated
masses using either X-rays.}
\begin{center}
\begin{tabular}{ccccc}
Name & Redshift & Nr. of galaxies & $M_{500}$ ($10^{14}$ M$_\odot $) & $\sigma _v$
  \\ \hline
PSZ2 G211.21+38.66  &  0.503 &   25   &    $7.0\pm 1.6$    &            $760\pm  150$   \\
PSZ2 G212.44+63.19  &  0.532 &   15   &    $4.2\pm 1.6$    &            $840\pm  270$   \\
PSZ2 G201.50-27.31  &  0.534 &   47   &    $9.3\pm 1.4$    &           $1430\pm  240$   \\
PSZ2 G094.56+51.03  &  0.541 &   55   &    $6.6\pm 1.6$    &           $1180\pm 180$   \\
PSZ2 G228.16+75.20  &  0.542 &   26   &    $11.0\pm 1.9$   &           $1130\pm 250$   \\
PSZ2 G111.61-45.71  &  0.547 &   30   &    $9.6\pm 1.4$    &           $700\pm  140$   \\
PSZ2 G183.90+42.99  &  0.559 &   21   &    $6.6\pm 1.6$    &           $1000\pm 260$   \\
PSZ2 G099.86+58.45  &  0.618 &   13   &    $7.1\pm 1.6$    &           $1000\pm  300$   \\
\end{tabular}
\end{center}
\label{Tab:clusters}
\end{table*}

In Fig. \ref{Fig:sigmav}, we see that the points for 178 clusters are close to the predictions
of virial theorem within a virial radius of $r_{200}$ for Newtonian gravity, or for MOND
with some parameters.
The best power-law weighted fit (taking into account both the errors of masses and velocities) is:
\begin{equation}
\sigma _{v,{\rm best\ fit\ data}}=(613\pm 22)
\left(\frac{M_{500}}{10^{14}\ {\rm M_\odot }}\right)^{0.230\pm 0.027}\ 
{\rm km\ s^{-1}} 
\end{equation}
The data present a correlation between $X=\ln \left(\frac{M_{500}}{10^{14}\ {\rm M_\odot }}\right)$ and $Y=\ln
\left(\frac{\sigma _v}{\rm km/s}\right)$, including the same weighting factors: 
\begin{equation}
C\pm \Delta C=\left(\frac{\langle X\,Y\rangle}{\langle X\rangle \langle Y\rangle}-1\right) \pm
\left( \frac{\sigma _X\sigma _Y}{\sqrt{N}\langle X\rangle \langle Y\rangle }\right)
=(9.3\pm 1.3)\times 10^{-3}
,\end{equation}
a correlation at 7.2$\sigma $ level. This sigma-level does not strictly correspong to a Gaussian distribution, but
practically indistinguishable from a Gaussian one. 
Perhaps the complement to one of the confidence level might be somewhat larger than the seven
sigmas Gaussian one ($\sim 10^{-12}$), but it may certainly be said that in the present case the null
hypotheses may be reject with at a confidence level larger than 99.99\%. 
The estimator of the correlation coefficient is a sum of some 178 terms. 
Each of these terms is the product of two
Gaussian variables, assuming that the errors of both mass and velocity
dispersion conform to Gaussian statistics. Therefore, each of these
terms are random variables following a Rayleigh distribution. The
Rayleigh distribution is somewhat more extended than a Gaussian, but
the central limit theorem assures that the sum of many variable following
that distribution, or any distribution with well defined mean and
variance, tends to a Gaussian. In fact, the sum of only four of them is
already quite close to a Gaussian, although not in the farthest positions of the
wings.

For the respective redshift ranges of $z$ (low: $<0.10$; intermedite: $0.10\le z<0.30$; high:$\ge 0.30$), we
get 
\begin{equation}
\sigma _{v,{\rm best\ fit\ data, z<0.10}}=(636\pm 28)
\left(\frac{M_{500}}{10^{14}\ {\rm M_\odot }}\right)^{0.203\pm 0.045}\ 
{\rm km\ s^{-1}} 
\end{equation}
\begin{equation}
\sigma _{v,{\rm best\ fit\ data, 0.10\le z<0.30}}=(550\pm 43)
\left(\frac{M_{500}}{10^{14}\ {\rm M_\odot }}\right)^{0.294\pm 0.050}\ 
{\rm km\ s^{-1}} 
\end{equation}
\begin{equation}
\sigma _{v,{\rm best\ fit\ data, z\ge 0.50}}=(690\pm 560)
\left(\frac{M_{500}}{10^{14}\ {\rm M_\odot }}\right)^{0.19\pm 0.39}\ 
{\rm km\ s^{-1}} 
\end{equation}

There is no significant difference in the
trend between low and high redshift clusters.

\section{Discussion and conclusions}

The relationship between velocity dispersion and
masses in clusters was known to work properly within Newton+DM
\citep[e.g.,][]{Evr08,Zha11,Mun13}, and also in the case of some modifications of gravity 
different from MOND without including DM \citep[e.g.,][]{Bro06}. However, they did not work in MOND \citep{San99,Poi05,Ett19}. 
We explored here the reason for this inconsistency and make
major improvements in the application of the virial theorem.
In particular our virial theorem analytical relationship of velocity dispersion in galaxies with given mass profiles
includes a pressure (surface) term, which, although its relevance is recognized in some literature \citep[e.g.,][]{The86,Car96,Car97,Gir98},
 is not usually considered in analytical calculations, although it is implicitly taken into account 
when carrying out numerical simulations.
We also applied an updated calibration of non-baryonic mass in the NFW profile, baryonic ratio
and baryonic profiles, either with an isothermal $\beta $ model or a \citet{Pat15} model.

Our results show that we can reconcile MOND with the virial theorem in clusters. This agreement
is obtained when: 1) the pressure term is taken into account in the virial theorem, which gives a 10--15\% higher velocity dispersion 
for MOND than for Newton+DM; 2) we
explore a range of possible parameters in the baryonic matter profile rather than adopting a fixed one. 
In particular for MOND we predict velocity dispersions equivalent to Newton+DM by
adopting a $\beta $ model with $\beta =0.55-0.70$, and core radii $r_c<0.30r_{500}$, which is in
agreement with the known data.
Lower concentration favours a higher MOND effect, so $x_{500}=\frac{r_c}{r_{500}}=0.3$ increases the dispersion of velocities by a factor
10--15\% with respect to $x_{500}=0.15$ for the same $\beta \approx 2/3$; decreasing $\beta $ with $x_{500}=0.15$ also decreases the concentration
and produces similar results. This last effect is easy to understand in MOND since
lower concentrations enhance the MOND effect because the galaxies spend a longer time during their orbits in the MOND regime
of low ($<a_0$) accelerations. Also the greater pressure term for MOND is due to a lower concentration of baryons than DM.
Calculations
without pressure and with default parameters ($x_{500}$ much lower than 0.3 and $\beta =2/3$) would give a $\sigma _v$  15--25\% lower
than Newton+DM. Given than the dynamical mass is proportional to $\sigma _v^3$, this means dynamical masses 40--60\% lower, and this
would explain the discrepances found in previous studies.

MOND in the regime of very low accelerations creates a field [`phantom mass';
\citep{Mil86,Mil09,Wu15,Lop21}]
which has an effect dynamically  similar to the presence of 
non-baryonic dark matter in Newtonian gravity. 
Here we observe that MOND fits the predictions of the virial theorem in rich clusters of galaxies,
which should not be surprising, given that the MOND phantom mass effect is equivalent to 
the non-baryonic dark matter. If some inconsistency arises, a revision of our knowledge of
the distribution of baryons would be needed because, with appropriate profiles and calibrations of 
$\frac{M_{\rm bar.}}{M_{\rm total}}$, there is always a mathematical solution able to mimic non-baryonic dark matter.

\section*{Acknowledgements}
Thanks are given to Rafael Barrena (Inst. Astrof. Canarias: IAC) for providing calculations of velocity dispersions
for NIKA2 clusters and search of information about their X-ray masses. Thanks are given to the anonymous referee
for helpful comments that have significantly improved the contents of the paper. Thanks are given to Terry Mahoney (language editor of IAC) for proof-reading of this text.
\\

{\bf DATA AVAILABILITY:} The data forming the basis of this article are available in the article or given references.
\\

\appendix

\section{Pressure (surface) term in the virial theorem}
\label{.pressure}

When the virial theorem is applied to a portion of  a stable gravitating system, it takes a
different form from when it is applied to the whole system. In the latter case, the pressure
term, which is present in general, cancels (asymptotically) and the familiar result of $2K+V=0$ [Eq. (\ref{virial})] holds, where
$K$ is the kinetic energy and $V$ is the potential energy.
However, the theorem is usually applied to the inner parts of a more extended system. This
is the case, for instance, when the entities forming the objects (i.e.\
galaxies in the case of cluster of galaxies) are increasingly more difficult to discriminate from
the interlopers. The relevance of this pressure (surface) term in standard gravity is known \citep[e.g.,][]{The86,Car96,Car97,Gir98}.

For the general form, we must use the general scalar virial theorem, which takes the
form
\begin{equation}
\sum _i\vec{F}_i\,\vec{r}_i=\sum _i m_i|\vec{v}_i|^2  
,\end{equation}
where $\vec{F}_i$ is the force acting on $i$-th particle and $\vec{r}_i$ its position vector. 
The sum $i$ extends to all particles. The right hand side of the equation 
is $2K$, while the left hand side is equal to $-V$ plus the `pressure term'. 
This last term appears because the force $\vec{F}_i$ acting on the i-th particle is due 
not only to the gravitational fields but also to pressure for particles at the boundary of the system.

The contribution
of this term is simply $3P_r\,V_r$, where $P_r$ is the radial pressure at the boundary of the virial radius $r_{\rm max}$,
assumed to be constant over it; and $V_r$ is the volume within it.  Assuming a stationary system with null average velocity, for which
\begin{equation}
\langle v_i^2\rangle =3\sigma _v^2
,\end{equation}
where $\sigma _v$ is the rms of the velocities along the line of sight.
Hence, after dividing by the total mass, the virial theorem reads

\begin{equation}
\sigma _{v,r<r_{\rm max}}^2=\sigma _{v,r<r_{\rm max}, P=0}^2
+\frac{P_r(r_{\rm max})\frac{4}{3}\pi r_{\rm max}^3}{M(r_{\rm max})}
,\end{equation}
\[
\sigma _{v,r<r_{\rm max}, P=0}^2=-\frac{V(r_{\rm max})}{3M(r_{\rm max})}.\]

For isotropic pressure,
\begin{equation}
P_r(r_{\rm max})=\sigma _{v,r=r_{\rm max}}^2\rho (r_{\rm max}).
\end{equation}
For a more general case, when the pressure is not isotropic,
\begin{equation}
P_r(r_{\rm max})=\frac{3}{3-2\beta _a}\sigma _{v,r=r_{\rm max}}^2\rho (r_{\rm max}).
,\end{equation}
where $\beta _a$ is the anisotropy parameter:
\begin{equation}
\beta _a=1-\frac{\sigma _\theta ^2}{\sigma _r^2}
,\end{equation}
\[3\sigma _v^2 =\sigma _r^2+2\sigma _\theta^2
,\]
and where we have used the fact that for spherical systems $\sigma _\theta ^2=\sigma _\phi ^2$ ($r$, $\theta $, $\phi $ denote radial,
declinational and azimuthal directions).

\section{The NFW profile}
\label{.NFW}

Assuming
a critical density of $\rho _c=8.5\times 10^{-27}$ kg m$^{-3}$  (for $H_0=67.4$ km s$^{-1}$ Mpc$^{-1}$), a virial radius equal to $r_{200}$ 
and concentration index $C$, 
the Navarro--Frenk--White (NFW, \citet{Nav97}) profile 
follows these relationships for the mass density $\rho (r)$ and the mass within the sphere of radius $r$ M(r):
\begin{equation}
\label{mr}
M(r)=M_{200}\times \frac{\left[\ln \left(1+\frac{C\,r}{r_{200}}\right)-\frac{C\,r}{r_{200}+C\,r}\right]}
{\left[\ln (1+C)-\frac{C}{1+C}\right]}
,\end{equation}\[
r_{200}=0.9834\ {\rm Mpc}\times \left(\frac{M_{200}}{10^{14}\ {\rm M_\odot }}\right)^{1/3}
,\]\[
r_{500}=0.7246\ {\rm Mpc}\times \left(\frac{M_{500}}{10^{14}\ {\rm M_\odot }}\right)^{1/3}
,\]\[
\rho(r)=\frac{\rho_ 0}{\frac{C\,r}{r_{200}}\left(1+\frac{C\,r}{r_{200}}\right)^2}
,\]\[
\rho _0=8.368\times 10^{12}\frac{C^3}{\left[\ln (1+C)-\frac{C}{1+C}\right]}\ {\rm M_\odot \, Mpc^{-3}}
,\]
where $M_{x}\equiv M(r_{x})$ and $r_x$ is the radius of the sphere for which the average 
density inside it is $x$ times the critical density $\rho _c$. That is, $x\rho _c=\frac{M(r_x)}{\frac{4}{3}\pi r_x^3}$.

\section{Profile for baryonic matter}
\label{.beta}

We assume
a critical density of baryonic matter $\rho _{cb}=\rho _c\frac{\Omega _b}{\Omega_m}$
with $\rho _c=8.5\times 10^{-27}$ kg m$^{-3}$ (for $H_0=67.4$ km s$^{-1}$ Mpc$^{-1}$); 
$\Omega _m=0.315$, $\Omega _b=0.0493$ \citep{Pla20}.
For a mass density profile $\rho _{\rm bar.}(r)=\rho _0J\left(\frac{r}{r_c}\right)$,
the mass within the sphere of radius $r$ is:
\begin{equation}
\label{mrb}
M_{bar}(r)=\frac{M_{bar.500}}
{I(r_{500})}I(r)
,\end{equation}\[
I(r)=\int _0^{r/r_c}dx\,x^2J(x)
,\]\[
r_{200}=1.357r_{500}\left( \frac{I(r_{200})}{I(r_{500})} \right)^{1/3}
,\]\[
r_{500}=1.345\ {\rm Mpc}\times \left(\frac{M_{bar.500}}{10^{14}\ {\rm M_\odot }}\right)^{1/3}
\]\[
\rho _0=\frac{M_{bar.500}}{4\pi r_c^3I(r_{500})}
.\]
where $M_{bar.500}\equiv M_{bar}(r_{500})$ and $r_x$ is the radius of the sphere for which 
the average density inside it is $x$ times the critical baryonic density $\rho _{cb}$. The value of $r_{200}$ is solved iteratively.

An isothermal $\beta $ model is usual in the description of gas in clusters of galaxies \citep{Cav76},
where $J(x)=\frac{1}{(1+x^2)^{1.5\beta }}$.
For $\beta =2/3$, which is usual the assumed value \citep{Arn09}, $I(r)$ has an
analytical solution: $I(r)[\beta =2/3]=\frac{r}{r_c}-\tan ^{-1}\left(\frac{r}{r_c}\right)$.
A value of a cluster core radius scale equal to $r_c\sim 0.25$ Mpc is expected \citep{Jon84}; although a dependence onthe mass
is also expected.

\end{document}